\documentclass[9pt,nofootinbib,superscriptaddress,longbibliography,twocolumn]{revtex4-2}
\usepackage{times}
\usepackage{graphicx}
\usepackage{dcolumn}
\usepackage{bm,bbm}
\usepackage{tabularx}
\usepackage{xcolor}
\usepackage{tcolorbox}
\usepackage{algorithm}
\usepackage{algpseudocode}
\usepackage{amsmath}
\usepackage{bbold}
\usepackage{braket}
\usepackage{nameref}
\usepackage[toc,page]{appendix}
\DeclareUnicodeCharacter{00A0}{ }
\DeclareUnicodeCharacter{2212}{-}

\usepackage[colorlinks,breaklinks,linkcolor={blue},citecolor={magenta},urlcolor={blue}]{hyperref}

\makeatletter
\newcommand\org@hypertarget{}
\let\org@hypertarget\hypertarget
\renewcommand\hypertarget[2]{%
  \Hy@raisedlink{\org@hypertarget{#1}{}}#2%
  }
\makeatother

\begin{document} 

\title{Are Brain-Computer Interfaces Feasible with Integrated Photonic Chips?}

\author{Vahid Salari}
\email{vsalari@bcamath.org}
\affiliation{BCAM - Basque Center for Applied Mathematics; Alameda de Mazarredo, 14,48009 Bilbao, Basque Country - Spain}
\affiliation{Quantum Biology Laboratory, Howard University, 2400 Sixth Street NW, Washington District of Columbia, 20059, United States of America}

\author{Serafim Rodrigues}
\email{srodrigues@bcamath.org}
\affiliation{BCAM - Basque Center for Applied Mathematics; Alameda de Mazarredo, 14,48009 Bilbao, Basque Country - Spain}

\author{Erhan Saglamyurek}
\affiliation{Department of Physics, University of Alberta, Edmonton, AB T6G 2E1, Canada}
\affiliation{Department of Physics and Astronomy, University of Calgary, Calgary, AB T2N 1N4 Canada}
\affiliation{Institute for Quantum Science and Technology, University of Calgary, Calgary, AB, T2N 1N4, Canada}

\author{Christoph Simon}
\affiliation{Department of Physics and Astronomy, University of Calgary, Calgary, AB T2N 1N4 Canada}
\affiliation{Institute for Quantum Science and Technology, University of Calgary, Calgary, AB, T2N 1N4, Canada}
\affiliation{Hotchkiss Brain Institute, University of Calgary, Calgary, AB, T2N 1N4, Canada}

\author{Daniel Oblak}
\email{doblak@ucalgary.ca}
\affiliation{Department of Physics and Astronomy, University of Calgary, Calgary, AB T2N 1N4 Canada}
\affiliation{Institute for Quantum Science and Technology, University of Calgary, Calgary, AB, T2N 1N4, Canada}

\begin{abstract}
\noindent The present paper examines the viability of a radically novel idea for brain-computer interface (BCI), which could lead to novel technological, experimental and clinical applications. BCIs are computer-based systems that enable either one-way or two-way communication between a living brain and an external machine. BCIs read-out brain signals and transduce them into task commands, which are performed by a machine. In closed-loop, the machine can stimulate the brain with appropriate signals. In recent years, it has been shown that there is some ultraweak light emission from neurons within or close to the visible and near-infrared parts of the optical spectrum. Such ultraweak photon emission (UPE) reflects the cellular (and body) oxidative status, and compelling pieces of evidence are beginning to emerge that UPE may well play an informational role in neuronal functions. In fact, several experiments point to a direct correlation between UPE intensity and neural activity, oxidative reactions, EEG activity, cerebral blood flow, cerebral energy metabolism, and release of glutamate. Here, we propose a novel skull implant BCI that uses UPE. We suggest that a photonic integrated chip installed on the interior surface of the skull may enable a new form of extraction of the relevant features from the UPE signals. In the current technology landscape, photonic technologies advance rapidly and poised to overtake many electrical technologies, due to their unique advantages, such as miniaturization, high speed, low thermal effects, and large integration capacity that allow for high yield, volume manufacturing, and lower cost. For our proposed BCI, we make some major conjectures, which need to be experimentally verified, and hence we discuss the controversial parts, feasibility of technology and limitations, and potential impact of this envisaged technology if successfully implemented in the future.
\end{abstract}

\date{\today}
\maketitle


\section{Introduction}
Brain-computer interface (BCI), or generally brain-machine interface (BMI), are computer (machine)-based systems that map brain signals into computer (machine) commands or actions. This mapping may involve intermediate analysis and processing. Moreover, a closed-loop BCI is also possible, whereby the brain is stimulated via relevant neuro-bio-signals. The most common brain signals used in BCI's are electromagnetic, that is, of classical/non-quantum origin. Herein, we turn attention to an exciting and emergent literature that reveals the brain also emits "photons", which are quanta of electromagnetic waves. The intensity of these emissions varies from a few photons to several hundred photons per second per square centimeter, mainly with spectral range of 200–800 nanometers \cite{Salari2015}. A caveat is that most single-photon sensitive detectors used in the experiments were only sensitive up to about 900~nanometers. Hence, observations with detector platrforms that are sensitive in the 900-1600~nanometer range, such as superconducting nanowire single-photon detectors (SNSPDs) \cite{marsili2013}, which also can be shaped as arrays \cite{wollman2019}, may reveal hidden obscured about the UPE light.

The body of evidence for ultraweak photon emission (UPE) is fast growing and is being independently observed by different scientific communities/labs. Due to infancy of the research field, many different terms are used to describe this phenomenon, including biophotons, ultraweak photon emission, ultraweak bioluminescence, self-bioluminescent emission, photoluminescence, delayed luminescence, ultraweak luminescence, spontaneous chemiluminescence, ultraweak glow, biochemiluminescence, metabolic chemiluminescence, dark photobiochemistry, etc. In this report we will henceforth adopt the term UPE. It has been evidenced that neurons and other living cells (e.g. in plants, animals, and humans) have spontaneous UPE \cite{Cifrareview} mediated via their metabolic reactions associated with physiological conditions. In 1967, it was first reported that electric pulses in neurons can induce weak photon emission (in the visible region of the EM spectrum) due to chemical reactions accompanying pulses, while a dead-neuron does not exhibit any photon emission \cite{Artemev}. In 1984 \cite{Imiazumi} and 1985 \cite{Suzuki} it was demonstrated experimentally that after the induction of hypoxia states in a rat brain, UPE increases. 
Isojima et. al \cite{Isojima} in 1995 showed that there is a correlation between the intensity of UPE and neural metabolic activity in the rat hippocampal slice. In 1997, Zhang et al. \cite{Zhang} revealed that the intensity of UPE from intact brains isolated from chick embryos was higher than the medium in which the brain was immersed. In 1999, Kobayashi et al.\cite{Kobayashi} detected spontaneous UPE in the rat’s cortex in vivo without adding any chemical agent or employing external excitation and found that the UPE correlates with the Electroencephalography (EEG) activity, cerebral blood flow and hyperoxia, and the addition of glutamate increases UPE, which is mainly originated from the energy metabolism of the inner mitochondrial respiratory chain through the production of reactive oxygen species (ROS). Kataoka et al.\cite{Kataoka} detected spontaneous UPE from cultured rat cerebellar granule neurons in the visible range and demonstrated that the UPE depends on the neuronal activity and cellular metabolism. Then, a fascinating experimental discovery by Sun et al. revealed that photons can be conducted along neuronal fibers. In 2011, Wang et al.\cite{Wang} show-cased in-vitro experimental evidences of spontaneous UPE and visible light induced UPE (delayed luminescence) from freshly isolated rat’s whole eye, lens, vitreous humor, and retina. Subsequently, in 2014 \cite{Tang2} Tang and Dai provided experimental evidence that the glutamate-induced UPE can be transmitted along the axons and in neuronal circuits in mouse.

These observations raise the following intriguing question: what are the underlying physiological processes that underpin UPE? Specifically, in the brain what are the associated neurophysiological processes? Although, a complete picture has not been provided, has been shown that the origin of UPE is in direct connection with the ROS. Moreover, its intensity has a direct correlation with thermal, chemical and mechanical stress, the mitochondrial respiratory chain, cell cycle, neural activity, EEG activity, cerebral blood flow, cerebral energy metabolism, and release of glutamate. Experiments also show that cells can absorb photons by photochemical processes and slowly release these photons as delayed luminescence \cite{Scordino}. Interestingly, it has been shown that delayed luminescence emitted from the biological samples provide valid and predictive information about the functional status of biological systems \cite{Niggli1, Niggli2, Musumeci}. All this opens novel exciting mathematical and physical questions at the interface of quantum biology. For example, if we consider UPE in the context of metabolism, then there has been efforts to propose quantum-metabolism \cite{Demetrius}. As known, biological systems are essentially isothermal and as such energy flow in living organisms is mediated by differences in the turnover time of various metabolic processes in the cell, which occur cyclically. The mean cycle time ($\tau$) of these metabolic processes (turnover of essentially redox reactions) are related to the metabolic rate ($g$), that is, the rate at which the organism transforms the free energy of nutrients into metabolic work. This is related to two coupled chains (electron-proton transport) of the ATP system in the mitochondria. In quantum-metabolism the main variables are metabolic rate, the entropy production rate and the mean cycle time. Then the fundamental unit of energy is given by $E(\tau) = g \tau$, where $g$ is related to the electron-proton transport. Noteworthy, this is in contrast, but has some correspondence to quantum thermodynamics, where we the thermal energy per molecule is given by $E = K_b T$, which relates specific heat, Gibbs–Boltzmann entropy and absolute temperature $T$. The difference is that biological systems work far from thermodynamic equilibrium hence in quantum-metabolism the variables depend on fluxes (rates of change of energetic values). On top of this, Albrecht-Buehler \cite{Albrecht-Buehler} hypothesized that the electron-proton transport releases photons ($E=h\nu$, where E is the photon energy, $h$ is plank constant and $\nu$ photons frequency). Other researchers have contemplated at why UPE displays wide variety of frequencies, with Popp suggesting that these are coherent and mediated by DNA, thus it may regulate life processes of an organism. However, the coherence idea of UPE is still under debate and it is yet unclear if UPE is just a byproduct in biological metabolism or it has some informational or functional role.\\
 
So far, UPE signals have only been studied in the context of basic science and has not been considered for experimental and clinical applications or novel technologies such as BCIs. The present article takes that first step forward and propose an implant BCI chip based on UPE. Since UPE is correlated to several sub-cellular, cellular and neural tissue processes, there is also the potential that it can be used as a novel technological probe/bio-marker for both normal brain function and pathological conditions. In the subsequent sections, we will first briefly review the traditional classical methods in BCI and then we will focus our discussion towards UPE detection and pattern recognition for the development of a novel UPE-based skull implant BCI.

\section{Classical Brain-Computer Interface Technology}

In traditional BCI techniques, different types of signal acquisition may be used, depending on the application. In the following, we briefly review four types of brain signals, their properties and the suitable machine interfaces.

\begin{itemize}
    \item {\bf Electroencephalography (EEG) Signals}\\
     EEG is the most employed method to detect electrical activity of the brain by use of small electrodes attached to the scalp \cite{EEG}. These signals are recorded by a machine for tracing both normal brain function and diagnosing pathological conditions (e.g. epilepsy). In stimulus (e.g. visual cue) induced EEG, there is positive deflection of voltage with a latency (delay between stimulus and response) of roughly 250 to 500 ms, which is called event-related potentials (ERP). Examples of such ERP is the so called  P300 formed at time 300ms, which is related to decision making. Indeed, cognitive impairment is often correlated with modifications in the P300 \cite{P300}. It is considered an endogenous potential, as its occurrence links not to a stimulus's physical attributes, but a person's reaction to it. More specifically, the P300 is thought to reflect processes involved in stimulus evaluation or categorization. The presence, magnitude, topography, and timing of this signal are often used as metrics of cognitive function in decision-making processes and hence used in BCIs. The P300 has several desirable qualities for pattern recognition. First, the waveform is consistently detectable and is elicited in response to precise stimuli. The P300 waveform can also be evoked in nearly all subjects with little variation in measurement techniques, which help simplify interface designs and permit greater usability. The speed at which an interface can operate depends on how detectable the signal is despite "noise." One negative characteristic of the P300 is that the waveform's amplitude requires averaging multiple recordings to isolate the signal. This and other post-recording processing steps determine the overall speed of a BCI interface \cite{P300BCI}. 
    
    \item {\bf Magnetoencephalography (MEG) signals}\\
     MEG is a functional neuroimaging technique monitoring brain activity via magnetic fields of electrical currents in the brain, using SQUIDs (superconducting quantum interference devices), which are very sensitive magnetometers operated in a cryogenic environment. Another type of magnetometer is spin exchange relaxation-free (SERF) magnetomere \cite{Hamalainen}, which can increase portability of MEG scanners, while it features sensitivity equivalent to that of SQUIDs. A typical SERF magnetometer is relatively small, and does not require bulky cooling system to operate. It has been demonstrated that MEG could work with a type of SERF, i.e. chip-scale atomic magnetometer (CSAM) \cite{Sander}, where its development can be used efficiently for BCI. Basically, MEG may provide signals with higher spatiotemporal resolution than EEG, and therefore useful for an increased BCI communication speed.
    
    \item {\bf Electrocorticography (ECoG) Signals}\\
    ECoG uses electrodes placed directly on the surface of the brain to record electrical activity from the cerebral cortex, i.e. an invasive technology that involves removing a part of the skull to expose the brain surface to enable the implant of an electrode grid on the surface of the brain, i.e. called craniotomy, which is a surgical procedure performed either under general anesthesia or under local anesthesia if patient interaction is required for functional cortical mapping. The spatial and temporal resolution of the resulting signal is higher and the signal to noise ratio (SNR) superior to those of EEG due to the closer proximity to neural activity. Thus, ECoG is a promising recording technique for use in BCI, especially for decoding imagined speech or music, in which users simply imagine words, sentences, or music that the BCI can directly interpret \cite{Shenoy}.

    \item {\bf Functional near-infrared spectroscopy (fNIRS) Signals}\\
     fNIRS is a noninvasive optical imaging technique that measures changes in hemoglobin (Hb) concentrations in the brain by means of the characteristic absorption spectra of Hb in the near-infrared (NIR) range \cite{Felixneuroimage}. fNIRS Tomography makes use of the fact that light penetrates up to several centimeters into biological tissue, i.e. a safe technique that is minimally minimally invasive and which relies on small, relatively inexpensive easy-to handle technology, and provides relatively low spatial resolution. The penetration range of light in tissue limits the size of the target tissue volume. fNIRS can be used in BCI for the restoration of movement capability for people with motor disabilities. fNIRS cannot afford high error rates for safety purposes, and must be fast enough to provide real-time control. Several fNIRS-BCI studies have tried to improve classification accuracies and information transfer rates \cite{Naseer}.
\end{itemize}

\section{Potential Application of UPE in BCI}
UPE is largely mediated by cellular metabolism and it is presently believed that it is merely a byproduct (i.e. epiphenomenon). A tempting question is whether it is possible (or not) to retrieve information from stochastic emission of UPE? In previous sections we already saw that there are different experimental reports on significant correlations between UPE emission and neuronal activity and associated metabolic processes \cite{Isojima, Kobayashi, Tang1, Kataoka}. Therefore, even if UPE is an epiphenomenon, its intensity can be a proxy for tracking the underlying neural information that dynamically changes under various conditions. Indeed, UPE seem to include information for monitoring physiological variations in a neuronal tissue. Note that for EEG signals we have a similar scenario. Indeed, EEG signals do not provide specific information about single neurons. Rather, it reflects a non-trivial summation of the synchronous activity of thousands of neurons and not that of a single neuron or dendrite. Thus, retrieving patterns as information from EEG is a data-science activity typically involving statistical comparisons between different brain states (e.g. normal and abnormal brain states).

Scholkmann \cite{Scholkmann1, Scholkmann2} hypothesized that UPE may be is used by neurosystems as an additional signal enabling cell-to-cell communication and coupling. Indeed, Sun et al.\cite{Sun} found that UPE can conduct along the neural fibers. It has been hypothesized based on numerical simulations that neurons (or myelinated axons) may act as optical fibers and, hence, may conduct light associated with UPE \cite{Kumar}, and through these waveguides UPE may even mediate long-range quantum entanglement in the brain \cite{Kumar, Zarkeshian}. These myelinated axons are tightly wrapped by the myelin sheath, which has a higher refractive index \cite{Antonov} than the inside of the axon and the interstitial fluid outside. Myelin is an insulating layer (sheath) around nerves, which is formed by two types of specialized glial cells, oligodendrocytes in the central nervous system (CNS) and Schwann cells in the peripheral nervous system \cite{Simons}. Muller glia cells have also been suggested to guide photons within mammalian eyes \cite{Agte, Franze, Reichenbach}. These observations suggest that UPE and bioelectronic activities are not independent biological phenomena in the nervous system, and their synergistic action may take on considerable function in neural (quantum) signal and information processes.

\subsection{UPE intensity from the surface of the human brain}
The UPE observed to date has been extremely weak. However, the true UPE intensity within neurons can be significantly higher than the one expected from the UPE measured a short distance away from the brain, as was done in all previous observations. Since photons are strongly scattered and absorbed in cellular or neural systems, the corresponding intensity of UPE within the organism or brain can even be two orders of magnitude higher \cite{Chwirot, Slawinski}. Based on the data from experiments with rat brain -- employing a 2D photon-counting tube with a photocathode featuring a minimum detectable radiant flux density of $9.9\times 10^{-17} W/{\text{cm}}^2$ under 1-s observation time -- the intensity of UPE has approximately $100$ $\frac{\text{counts}}{\text{sec.cm}^2}$ from the cortex surface \cite{Kobayashi1, Kobayashi, Adamo, Imaizumi}. Moreover, the limited quantum efficiency of the detector may impede the detection of UPE due to the limited SNR. Regarding the human brain, the neuronal density in V1 in visual cortex is $60\times 10^6\frac{\text{Neurons}}{\text{cm}^3}$ in postmortem human brains \cite{Pakkenberg}. It should be noted that postmortem studies use fixatives, which lead to shrinking of the tissue. The result is that the cell density is overestimated while the volume of the extracellular space is underestimated. The reported number can be used only as an absolute best-case scenario for the interface. The V1 thickness is about 0.2 cm, and V1 surface area of one hemisphere is about 26 cm$^2$ in adult humans. At least, 10$^6$ neurons in object-related areas and 30$\times 10^6$ neurons in the entire visual cortex are activated by a single-object image \cite{Levy}. Based on a rough estimation, about $10^6$ free radicals can be produced by each brain cell per second \cite{JPPB2010}, which yields $10^6\times 10^6=10^{12}$ free radicals produced by human visual neurons per second in V1 of one hemisphere during perception of a single-object image. Since UPE mainly originates from free radicals, the actual UPE intensity inside neuronal cells is expected to be considerably higher than the intensity measured by a detector outside (e.g. 100 counts/(s.cm$^2$)). If the quantum efficiency of an ideal photodetector is close to 100\% we conjecture that it may measure the UPE intensity at the cortex surface at least on the order of 1000 counts/(sec.cm$^2$) for an object visualization.

\section{Skull-implant Setup for the UPE-based BCI}
\begin{figure*}
    \centering
    \includegraphics[scale=0.35]{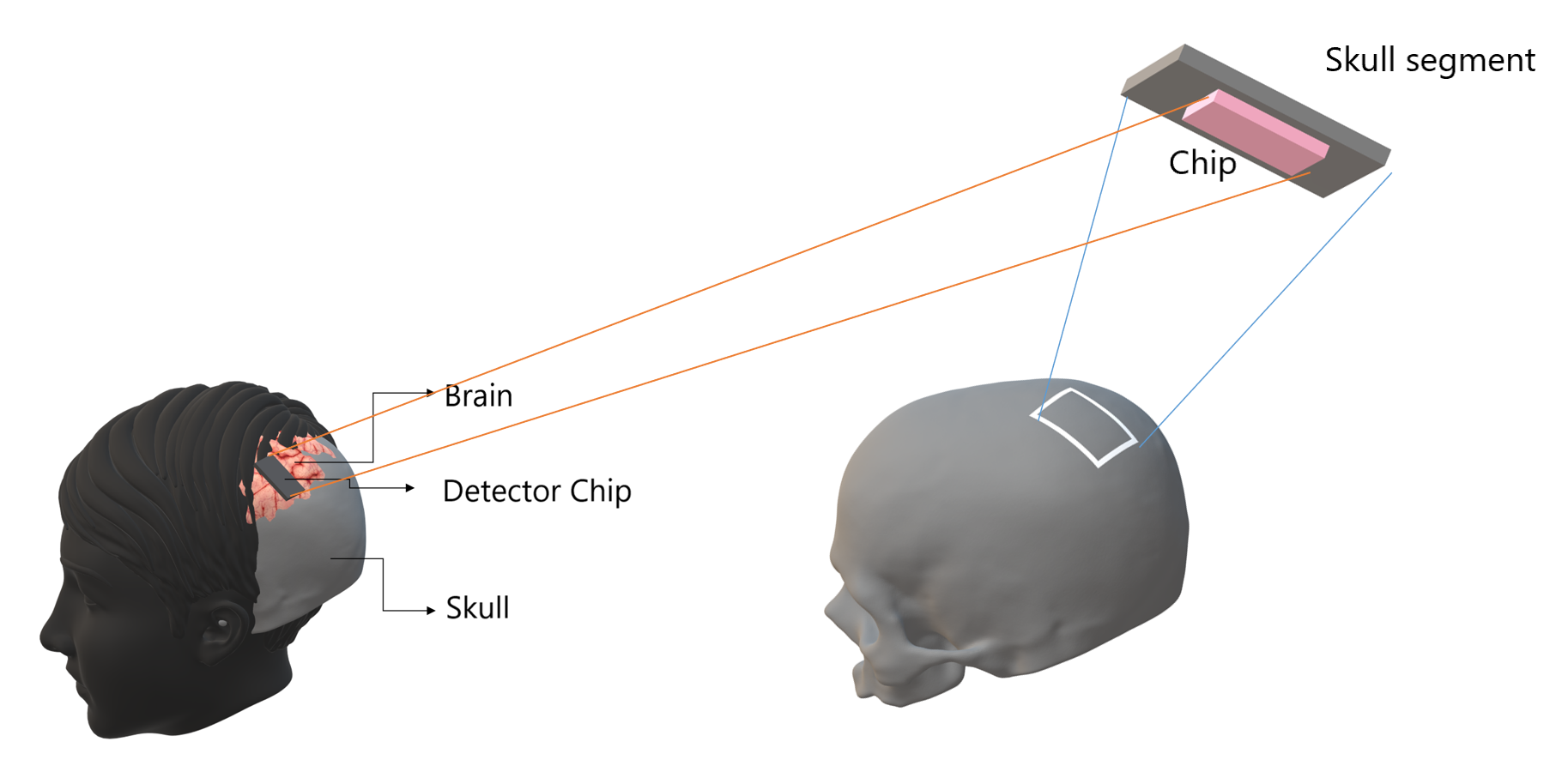}
    \caption{A detector chip can be installed on the interior surface of the skull without touching the brain tissue, i.e. non-invasive. The environment of the closed skull in the head is sufficiently dark and therefore it is a suitable environment for the detection of UPE with the installed chip. The intensity of UPE is stronger close to the surface of the brain, which can be captured by chip on the skull.}
    \label{fig:1}
\end{figure*}

\begin{figure*}
\centering
\includegraphics[width=18cm]{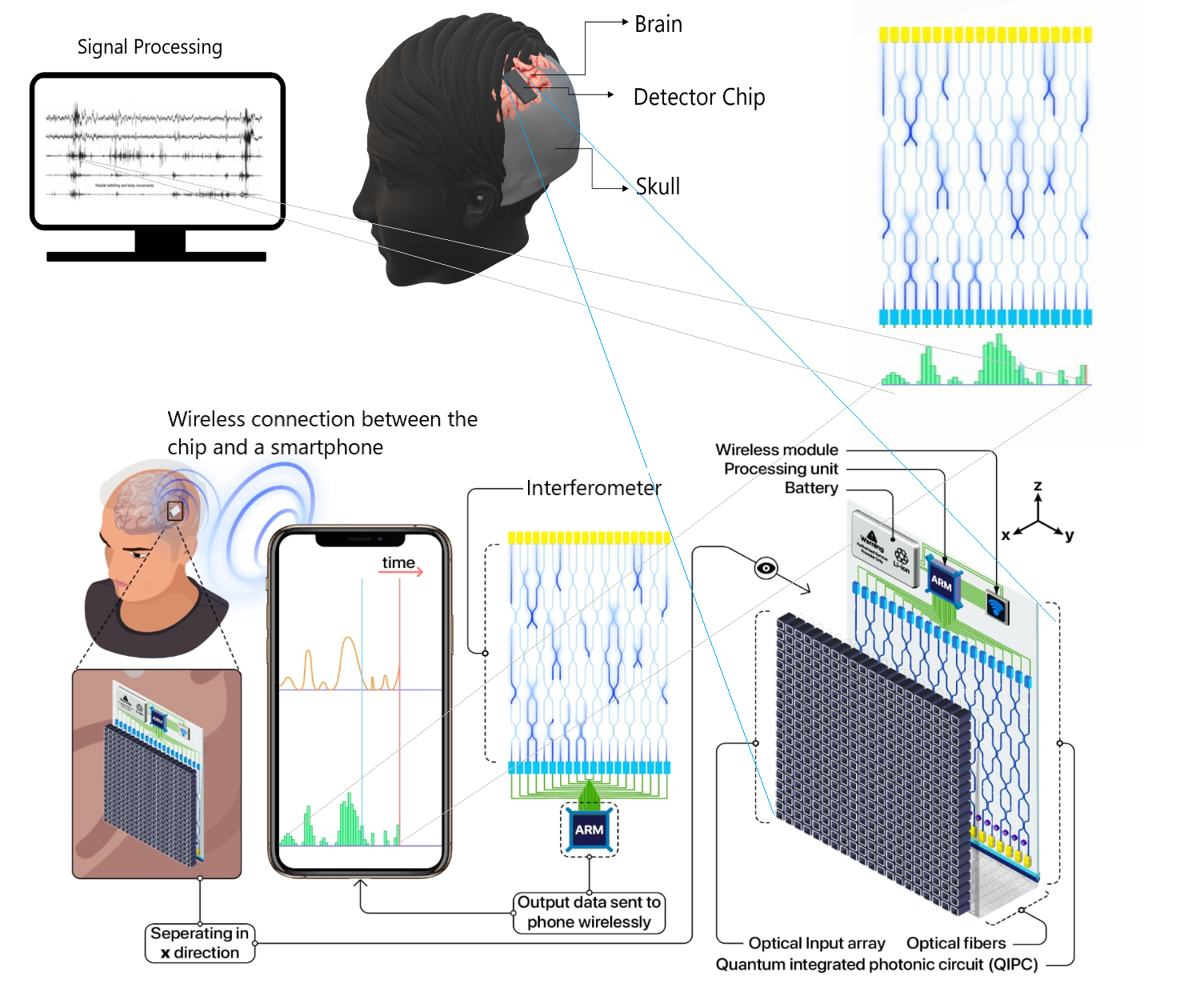}
\caption{ In a BCI proposal, an optical chip is implanted on the interior surface of the skull. A few number of UPE photons interfere in an a photonic chip and the results are detected as different single photon distributions at detectors versus time. This results are communicated via wireless signals from the detector part of the chip to a receiver (e.g. smartphone or a computer).}
\label{fig:2}
\end{figure*}

We now provide the complete design specification of a radically novel skull-implant that can facilitate a UPE-based BCI (see Fig. \ref{fig:1}). The envisaged BCI is not aimed for deep brain implants (although possible) but rather for intracranial brain surface implant (i.e. minimally invasive). The environment of a closed skull (after surgical implantation) is sufficiently dark and, therefore, a suitable environment for the detection of UPE signals. Once the UPE signals are detected, they are wirelessly relayed to a machine, computer, or smartphone. We also envisage alternative designs with closed-loop signals (photons) for modulating the metabolic processes of a neural tissue. However, herein we will only consider the read-out of UPE signals. The center-piece of the envisaged technology is the UPE-based integrated chip, which we will discuss at length in the subsequent subsections. The integrated photonic chip is assembled from different component parts; specifically, a receiver optical plane, optical fibers, a photonic interferometery circuit, a complementary metal-oxide-semiconductor (CMOS) detector array, a battery, and a wireless system (see Fig. \ref{fig:2}). The use of the implantable CMOS image sensor has been described in recent years especially for optogenetic imaging \cite{CMOSImaging}. 

The UPE photons first enter a receiver optical plane (ROP) on the chip, which is essentially a photo-receiver array made up of optical fibers, of size of $N\times N$, $N$ is the number of pixels (or fibers) in each row or column and each pixel is indeed an optical fiber that couples into a waveguide on the chip, using grating couplers \cite{Cheng2020}. Alternatively, the UPE light can be directly coupled to waveguides created by femto-second laser-writing and since these can be patterned at different depths in the chip \cite{nolte2003}, they can directly facilitate serialization step. Subsequently, the $N\times N$ pixels are serialised into a 1D vector (where $N'=N\times N$ is the number of optical fibers connected to the waveguides in the optical interferometer with $N'$ input ports in a series and linear 1D form, and therefore $N'$ CMOS pixels in a single row as the output port on the PIC). In fact, the received photons on ROP are guided to the optical interferometer via optical fibers. The advantage of an optical interferometer is that it may discriminate the emission patterns of photons. We estimate that UPE intensity ranges 10-1000 counts per second per each cm$^2$ of the whole array, depending on how active a neuron or neural tissue is at a given time instant. In fact, we expect that similar and non-similar UPE emission (in wavelength) generate different detection distributions, where interference will occur between photons with similar wavelength (i.e. emanating from the same-type neural processes). Thus, the detection distributions for  similar-wavelength photons will be closer to an optical interference pattern, which is uniquely determined by the wavelength of these interfering photons. In this regard, one of the concerns may arise from the fact that UPE emission over a broad range of wavelengths can lead to  the observation of different patterns at the same time, rendering an ambiguous combination of several independent patterns. Such complexity may bring disadvantages over the direct detection (i.e., no interferometer), or even could cause wrong interpretations. This potential problem can be alleviated by classifying those different wavelength patterns, again with pattern recognition techniques in machine learning, such as principal component analysis (PCA)\cite{PCA}, which allows distinguishing the differences in an ensemble of patterns, and identifying each pattern according to the respective wavelength, after many sets of training data. The optical interferometer photons are then converted into electrical signal via the CMOS array (see Fig. \ref{fig:5} for details). Finally, these signals are wirelessly linked to a smartphone or computer for pattern recognition/extraction. Noteworthy, since the number of detected photons is relatively low and because the data acquisition is in real time, the recognition of patterns should be done via machine learning protocols, e.g. convolutional neural networks (CCN), which is a powerful tool for 2D pattern recognition. We subsequently discuss in more detail each component part of the UPE-based electronic chip.

\subsection{On-Chip Photonic Integrated Circuits}
 We base our proposed technology on photonic integrated circuits (PICs) \cite{IPC}. These are chip that contains photonic components that operate with light (photons), where photons pass through optical components such as waveguides (equivalent to a resistor or electrical wire in an electronic chip). With electronic integrated circuits arriving at the end of their integration capacity, PICs have the potential to be the preferred technology. Nowadays, photonic platforms present several advantages for quantum information protocols enabling long coherence times, full connectivity, scalability, and operation in room temperature. Different photonic degrees of freedom, including polarization, spectral, spatial, and temporal modes can be used to encode information, providing different experimental resources for a wide variety of quantum information tasks. 

For our application we consider a PIC containing an optical interferometer. A linear interferometer can be fabricated through silica-on-silicon or laser-written integrated interferometers, or electrically and optically interfaced optical chips \cite{PIC1, PIC2, PIC3}, which makes a simple processor reducing the amount of physical resources needed for implementation.

\subsection{Photons Statistics and Distributions}

In the context of optics, coherence is a property of light. In a simplified picture, coherence is the ability of light to make interference, e.g. in the double-slit interference experiment light can create interference patterns (bright and dark bands) for both a wave (classical) and photon (quantum) picture.

Thus, coherence of light can be both of a classical and quantum character. For example, thermal states of light can be described in the classical and the quantum framework, while other states, such as squeezed states, can only be described in the quantum framework. One of the essential conditions to show the coherence property of light is for its intensity/photon-number distribution to be a Poisson distribution. However, this condition is not sufficient to conclude that the light is certainly coherent. Other types of sources may yield a Poisson distribution, too, e.g. shot noise and dark noise. In the following paragraphs we will introduce a couple of photon-number distributions in order to demonstrate, how this measure provides insight into the nature of the UPE light being emitted.

The photocount statistics of coherent light is a Poisson distribution \cite{Michalcoherence}
\begin{equation}
    P_n(t, T)=\frac{\langle n\rangle ^n}{n!}e^{-\langle n\rangle}
\end{equation}
where $\langle n\rangle$ is the average number of photons measured
between time t and time $t+T$. The variance of Poisson distribution is equal to its mean $\langle (\Delta n)^2\rangle=\langle n\rangle$. The deviation of the photon-number distribution from the Poisson distribution is measured by the Fano factor $F$ such that $\langle (\Delta n)^2\rangle=\langle n\rangle F$, or by the Mandel parameter $Q = F-1$. A photocount statistics is said to be super-Poissonian if $F>1$ and $Q>0$, and sub-Poissonian (and therefore non-classical) if $F<1$ and $Q<0$. Hence, the shift from a Poisson distribution is a sign of non-classical (quantum) characteristics of the light \cite{Michalcoherence} while a Poisson distribution is a sign of classicality.\\
The photocount statistics of a thermal source with M modes is approximated by the expression
\begin{equation}
    P_n(t, T, M)=\frac{(n+M-1)!}{n!(M-1)!})(1+\frac{M}{\langle n\rangle})^{-n}(1+\frac{\langle n\rangle}{M})^{-M}
\end{equation}
where $\langle n\rangle$ is the average number of photons and $M$ is the number of field modes \cite{Michalcoherence}. An important characteristic of these states is the relation between the variance and the mean $\langle (\Delta n)^2\rangle=\langle n\rangle +\frac{\langle n\rangle ^2}{M}$. The coefficient M is generally very large for chaotic sources. So that the relation between the variance and the mean is close to that of a coherent state, i.e., for large $M$,
$P_n(t, T, M)$ approaches a Poisson distribution (see Fig. 3). In relation to UPE, it is important to know whether the photocount statistics can distinguish between the coherent and thermal emissions, because photocount statistics of thermal light becomes equal to that of a coherent state when the number of modes $M$ is large. Since the photocount statistics are not able to discriminate between a coherent and a thermal state with many modes. 

Another type of emission is super-radiance, which is the coherent emission of light by several sources, and its main characteristic is the fact that the intensity of the emitted light can vary with the square of the number of sources because they can emit in phase. The photocount statistics of super-radiant emission is sub-Poissonian \cite{Michalcoherence}, and the photon state of a super-radiant system is generally not a coherent state.\\

\begin{figure*}
    \centering
    \includegraphics[scale=0.25]{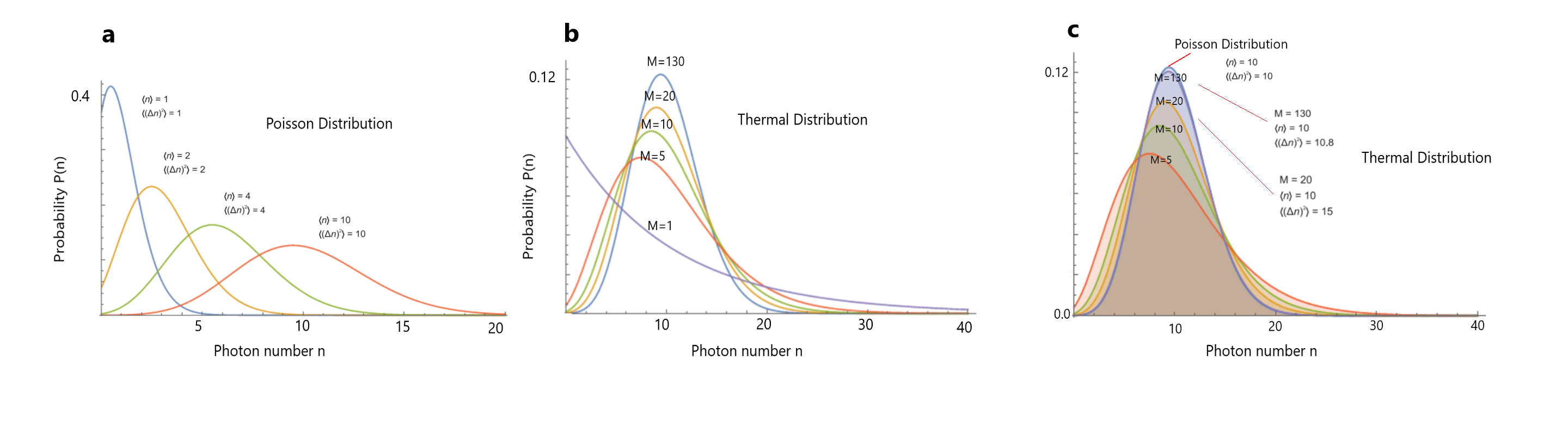}
    \caption{a) Poisson distribution for four different average values of photon counts $\langle n\rangle$. b) Demonstration of thermal field photocount distribution for different number of thermal modes for the average number of 10 photons. c) Thermal field photocount distribution (with similar $\langle n\rangle$) approaches Poisson distribution for a large number of modes $M$.}
    \label{fig:my_label}
\end{figure*}

\subsubsection{Photon Detection with Interference}
The photons collected onto our chip will then be propagated through a PIC featuring several interference paths and other components. The model of the effect of the PIC on the incident photons aims to predict the probability distribution of photons at the detector following their propagation and interference in a linear interferometer. The experimental setup only requires photodetectors and linear optical elements, i.e. beam splitters and phase shifters. Suppose the chip is injected with an input state of single photons of UPE, $|S\rangle=|s_1, s_2, ..., s_N'\rangle$ where $s_k$ are the number of UPE emitted photons in the $k$-th mode and injected into the chip. The output state of the chip can be written as $|O\rangle=|x_1, x_2, ..., x_N'\rangle$. For the sake of simplicity, suppose that there are four outputs on the chip. Therefore, probabilities of output detection for $N=1$ input photons in case there is no dissipation in the circuit are $P_{|1000\rangle}$, $P_{|0100\rangle}$, $P_{|0010\rangle}$, $P_{|0001\rangle}$, and for $N=2$ input photons the probabilities at the output are
$P_{|1001\rangle}$, $P_{|1010\rangle}$, $P_{|1100\rangle}$, $P_{|0110\rangle}$, 
$P_{|0011\rangle}$, $P_{|0101\rangle}$, $P_{|2000\rangle}$, $P_{|0200\rangle}$, $P_{|0020\rangle}$, $P_{|0002\rangle}$. Now, we consider a general case for $N'$ outputs. The signal processing and the interpretation of the signals require machine learning techniques. As the signal aqcuisition is performed through an interferometer, different interference patterns may form. We suggest a pattern recognition approach via convolutional neural networks (CNN) \cite{Fukushima} for an efficient interpretation of output signals on the photonic interferometer chip. Here, the conjecture is that a synchronous activity in a specific region of cortex makes synchronous similar metabolism with similar chemical reactions producing similar ROS byproducts simultaneously, and therefore the probability of detection of similar photons (even with a low probability of interference in the interferometer) during a specific brain activity is higher than the normal state with stochastic photon emissions. Discrimination between the interference pattern of active and normal states will be non-trivial but tractable via machine learning. This conjecture is expected to be reasonable based on highly synchronized brain activities for different specific cognitive tasks. In fact, the photonic chip continuously produces data under normal and active states of the brain. The patterns can be recognized by studying the data and classifications via discrimination between the signals of normal and active states. In such a state, both supervised and unsupervised learning can be performed on software. This can be an advantage of the method.  

\begin{figure*}
    \centering
    \includegraphics[scale=0.27]{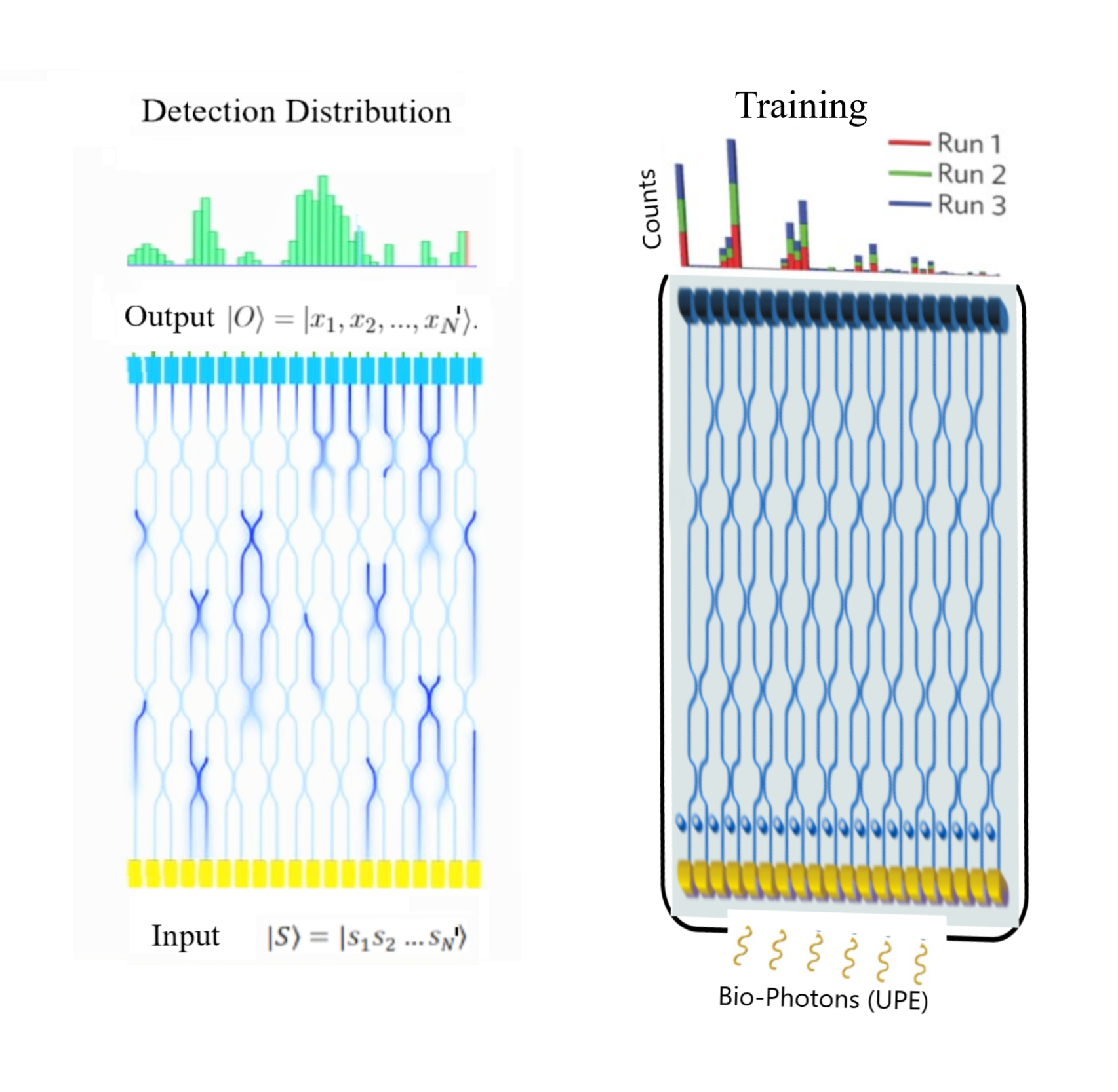}
    \caption{On-chip optical interferometer with N' inputs and N' outputs. The output patterns can be processed for feature extraction via machine learning techniques. It is expected that for each cognitive task or decision making, a similar pattern (in average) forms after many runs under training for specific tasks. The features of the average pattern can be recognized by deep learning methods, or specifically by convolutional neural networks (CNN) on a software.}
    \label{fig:3}
\end{figure*}

\begin{figure*}
    \centering
    \includegraphics[scale=0.30]{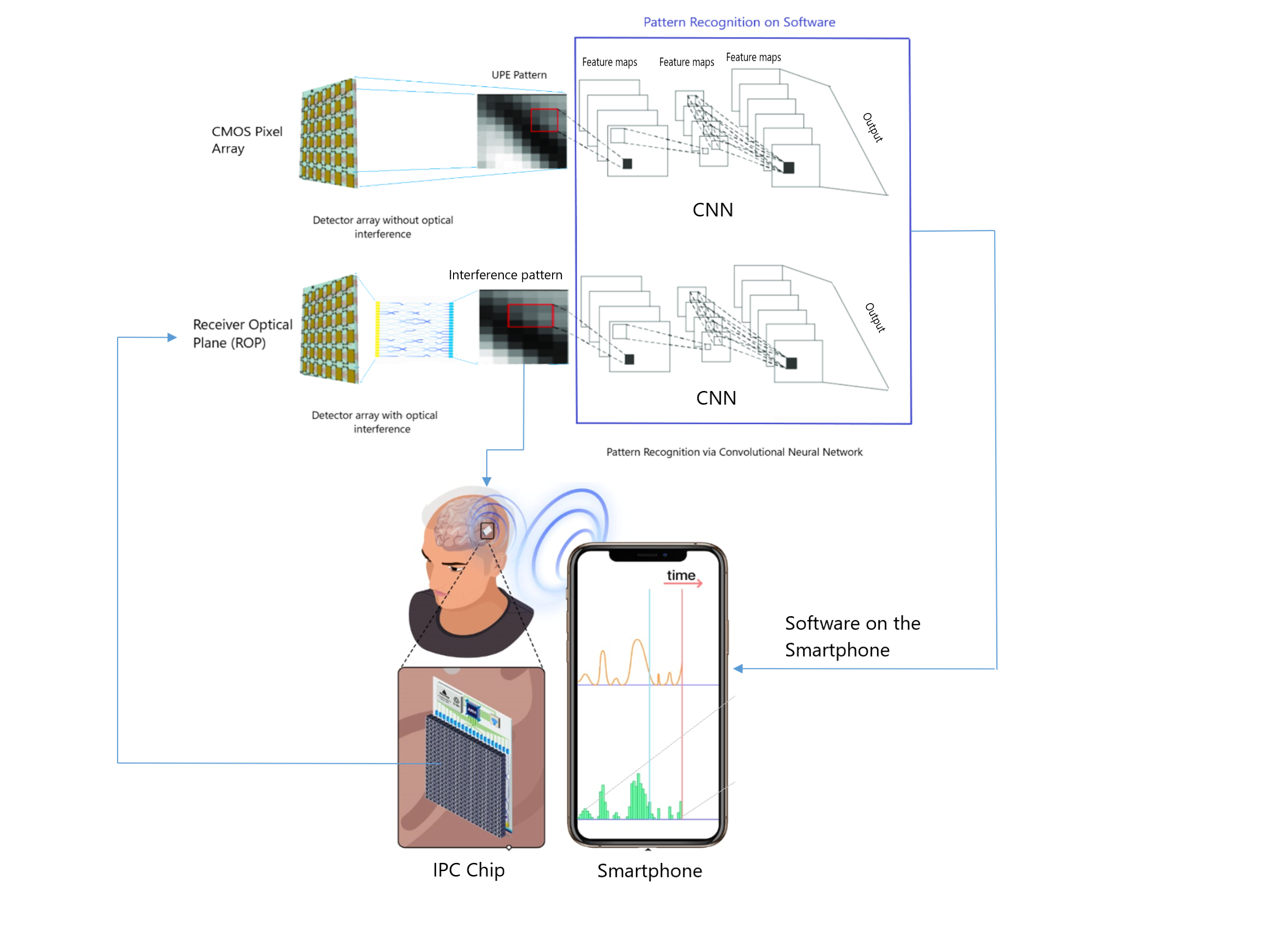}
    \caption{Feature extraction and pattern recognition of detected UPE by a chip composed of CMOS array via convolutional neural network (CNN) on a software installed on a computer, machine, or smartphone, top) direct UPE detection without optical interferometer, and bottom) UPE detection after the interferometer. The existence of optical interferometer is to discriminate UPE wavelengths, since interference of similar photons (in wavelength) make a different pattern with non-similar photons. One of the advantages of such an interferometer is to have a simple "spectrometry" over similar wavelengths. However, an ensemble of wavelengths may make different patterns at the same time and obscure the interference patterns which may not make advantage over a direct detection, but one can classify those ensemble patterns with pattern recognition techniques such as principal component analysis (PCA), which can find the differences between different patterns in the overlapped patterns, and classify each pattern for the relevant wavelength after many sets of training data. The direct detection of UPE by CMOS array and indirect detection after an optical interferometer both can be used for UPE data acquisition.}
    \label{fig:4}
\end{figure*}

The idea of using UPE signals for BCI applications still remains at the level of conjecture, relying on a mere fact that UPE shows correlations with some brain activities. Therefore, from a BCI point of view, such correlations are very important because for almost all types of brain signals for BCI applications, it is hardly possible to extract specific information from the signals directly. With an analysis of signals over thousands of training trials it will be possible to obtain an average pattern with specific features (for feature extraction) that finally make it easy for a specific algorithm to recognize the pattern in the next acquisition signals directly. Here, we suggest using a machine-learning algorithm to discriminate variations and extraction of features by enhancement of training data. A deep-learning algorithm becomes stronger in learning with increasing the training data to a specific level. This is a benefit for an implanted chip since it is always creating thousands of patterns easily to be processed by software on a computer or a smartphone. There is no need to perform separate experiments each time for training. Therefore, a deep-learning algorithm can learn how to understand features from UPE signals and interpret them according to the relevant cognitive task. Thus, data analysis of the output UPE signals of the chip can be performed via machine learning in general and deep learning specifically. For instance, a possibility is via deep learning method called CNN technique, which enables high-resolution pattern recognition. Since CNN are ideal for 2D imaging processing, then the UPE signals detected at the receiver optical plane pixel-array can be readily adapted for CNN (see Fig. \ref{fig:4}). The pattern analysis can be enhanced depending on the details our architecture. CNN error minimization methods are used to optimize convolutional networks in order to implement quite powerful pattern transformations. This is very useful when the input is spatially or temporally distributed. The first layer of a CNN generally implements nonlinear template-matching at a relatively fine spatial resolution, extracting basic features of the data. Subsequent layers learn to recognize particular spatial combinations of previous features, generating ’patterns of patterns’ in a hierarchical manner. If down-sampling is implemented, subsequent layers perform pattern recognition at progressively larger spatial scales, with lower resolution. A CNN with several down-sampling layers enables processing of large spatial arrays, with relatively few free weights. As we discussed before, an ensemble of wavelengths may make different patterns at the same time and obscure the interference patterns, where a PCA algorithm \cite{PCA} can find the differences between different patterns in the overlapped patterns, and classify each pattern for the relevant wavelength after many sets of training data. 

\subsection{Implementation Feasibility}
\begin{figure*}
    \centering
    \includegraphics[scale=0.20]{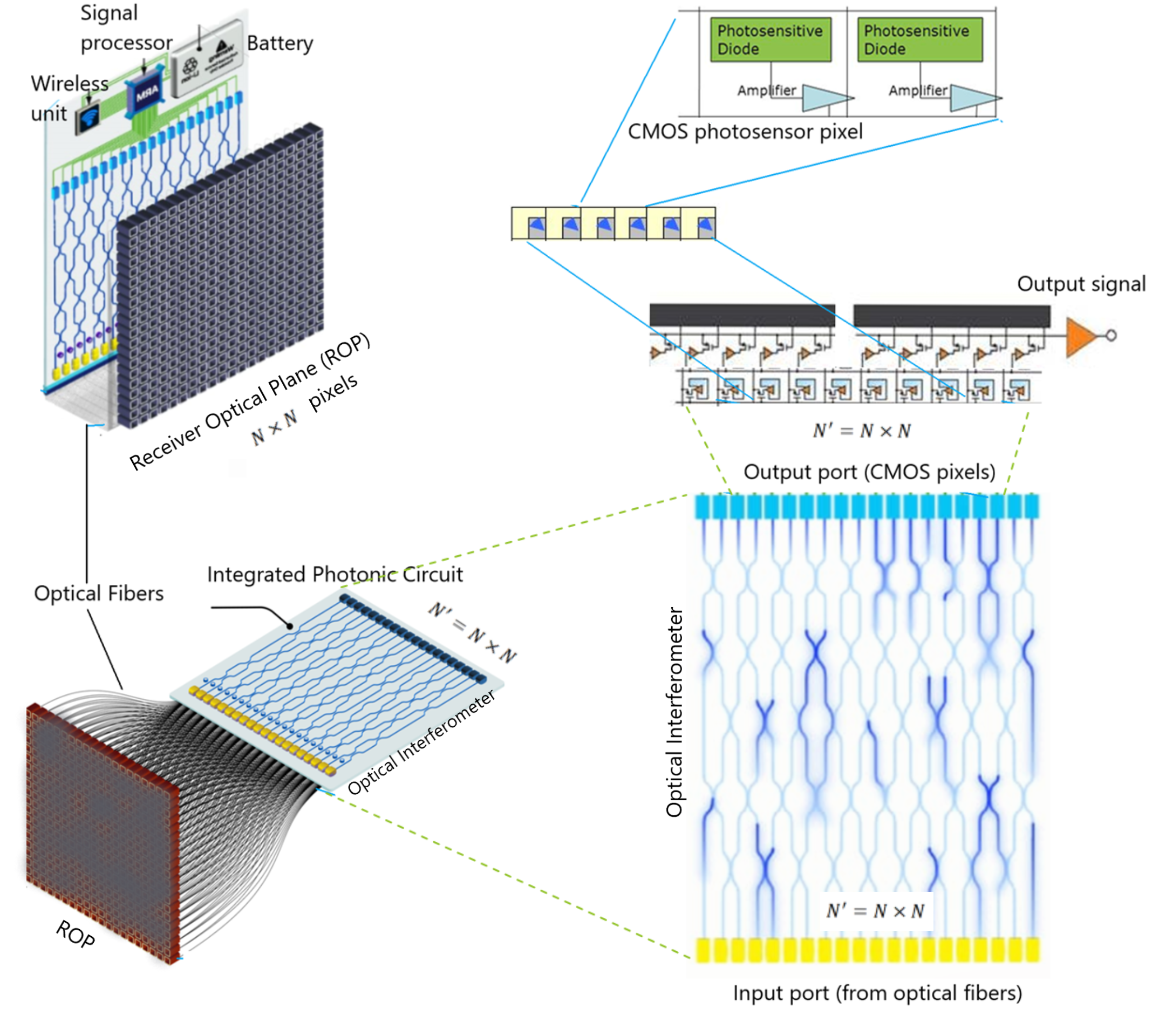}
    \caption{A typical on-chip UPE detector can be built from an array of optical fibers connected to an integrated photonic circuit which has an output gate composed of CMOS photosensor array.}
    \label{fig:5}
\end{figure*}

We now discuss the feasibility of fabricating all elements of our envisaged skull-implant UPE-based BCI (to be followed with Fig. \ref{fig:5}).

\subsubsection{Chip Ingredients}
The design and fabrication of PICs is a mature technology, which is realized on a variety of material platforms, which are tailored to the needs and requirements of the the application at hand. Available, platforms for lithography-based fabrication include Silicon photonics (Silicon on Insulator (220nm and 3 $\mu$m SOI), Si based Silica on Silicon (SiO$_2$, also known as PLC) and Silicon Nitride (SiN and TriPleX), III-V photonics such as Indium Phosphide (InP), Gallium Arsenide (GaAs) and derivatives, and finally Lithium Niobate (LiNbO$_3$) and other more exotic materials \cite{Chip1, Chip2, Chip3, Chip4, Chip5, IPC2, Chip9, Chip10}. It should be noted that the SIO platform is not a suitable candidate for the UPE in the visible spectrum as the relatiely small band-gap of silicon renders it completely opaque below a wavelength of about 1000~nm. SiN, which, on the other hand, is transparent in the visible wavelength-range and features compatibility with CMOS technology \cite{Chip6}, appears to be a strong candidate as a PIC platform for our proposed BCI. As an alternative to the lithography-based PIC, femto-second laser-written waveguides (FLWs) in SiO$_2$ (glass) have in recent years been used to successfully implement advanced PICs \cite{Chip7, Chip8}. The unique advantage of FLWs is that the ability to define waveguides in three dimensions, i.e., including at different depths in the chip. This allows more complex routing, such as the crossing of waveguides \cite{Chip8}.

Choosing the right technology will be the starting point for having a successful integrated chip. By integrating all devices into a single chip, complex assembly,
alignment and stabilization processes are avoided, and packaging and testing are greatly simplified. Moreover, it is the only way to scale up complexity when moving over 20-30 components into a single package. The selection of the integration material will then determine the capabilities and limitations for the technology platform, making some of them more appropriate for certain applications than others. This is thus a critical choice and needs to be carefully evaluated.

\subsection{Noise and Loss in the PIC}
Design of an PIC, testing and packaging from the beginning should be done carefully. The steps are device level (optical, thermal, and material simulations), circuit level (virtual lab to test performance), system level (PIC connected to a CMOS array), layout level (generate the design intent), verification, simulation of each process step, fabrication, and finally packaging. Moreover, a software should be designed to process the detected signals. Here, we would like to estimate the noise magnitude in the optical section of the PIC. The optical section is composed of receiver optical plane (ROP), optical fibers (OF), and optical interferometer (OI). 

\subsubsection{Noise and Loss in the receiver optical plane}
First, we note that blackbody radiation is not a significant source of photons in the visible wavelength range at body temperature. On our BCI, photons are directly coupled to the ROP's fibers that are very close (approximately in contact) with the cortex, thereby leading to a minimal coupling loss. In terms of noise, shot noise (also known as "quantum noise"\cite{Qnoise} or "photon noise") is the most important contribution in the ROP. It describes the fluctuations of the number of photons received due to their occurrence independent of each other. Optical detection is said to be "photon noise limited" as only the shot noise remains. Just as with other forms of shot noise, the fluctuations in a photo-current due to shot noise scale as the square-root of the average intensity:\\

\begin{equation*}
\text{SN}:= \mid \sqrt{(n-\langle n\rangle)^2}\mid
\end{equation*}

\subsubsection{Loss in optical fibers}
The intensity of photons will become lower when travelling through the core of fiber optic. Thus, the signal strength becomes weaker. This loss of light power is generally called fiber optic loss or attenuation. This decrease in power level is described in dB. There are two types of loss in optical fibers known as intrinsic fiber core attenuation (mainly due to light absorption and scattering) and extrinsic fiber attenuation due to bending loss as well as splicing (or coupling) loss between the fibers and chip. Given that the length of the fibers are to be in centimeter scale, the former  will be negligible. However, bending and splicing/coupling loss can be significant depending on the process of binding the fibers to the photonic chip. For example, based on subwavelength gratings, it has been shown that it is possible to couple broadband light with very low coupling losses. Guiding of visible light in the wavelength range of 550–650 nm with losses down to 6 dB/cm is feasible using silicon gratings (having absorption of 13,000 dB/cm at this wavelength), which are fabricated with standard silicon photonics technology. This approach allows one to overcome traditional limits of the various established photonics technology platforms with respect to their suitable spectral range \cite{LSA2021}.

\begin{table*}
\begin{tabular}{ |p{5cm}||p{2.5cm}|p{2.5cm}|p{2.5cm}| |p{2.5cm}| }
 \hline
 \multicolumn{5}{|c|}{Loss vs wavelength for various chip platforms}\\
 \hline
 \textbf{Loss (dB/cm)} & 300-400 nm  &400-500 nm & 500-600 nm &600-700 nm\\
 \hline
 Aluminum nitride (AlN)  & \ 40-50   &40-50&   30-40  &20-30\\
 Alumina (Al$_2$O$_3$) &$\sim3$  & 2   &1  &$<1$\\
 Tantalum pentoxide (Ta$_2$O$_3$) &N/A &$\sim{4}$ & $\sim2$ &$<1$\\
 Silicon-nitride (Si$_{3}$N$_{4}$)   &N/A &5-20 &$<1$ &$<1$\\
 Lithium niobate (LiNbO$_{3}$)   & N/A & N/A & N/A &$\sim 0.06$\\
 Femto-second laser-written waveguides in glass (SiO$_{2}$)   & N/A & N/A & N/A &$\sim 0.2$\\
  \hline
\end{tabular}
\caption{Data adapted from APL Photonics 5, 020903 (2020); Optica 6(3), 380-384 (2019); and Optics Express 14(11), 4826-4834 (2006)}
\label{tab}
\end{table*}

\subsubsection{Noise and Loss in optical interferometer}
The main elements of an optical interferometer on a photonic chip are couplers and optical modulators. There are different types of optical modulators such as
MEMS, liquid crystal on silicon (LCOS), electro-optic LiNbO$_{3}$ waveguide, III-IV semiconductor optical amplifier (SOA), Mach-Zehnder interferometer (MZI), and micro-ring resonator (MRR) \cite{CMOSSensors}. Compared with the above technologies, the silicon photonic modulators based on silicon-on-insulator
(SOI) platform attract more attention because of high device density, whose volume is 1/1000 of silicon dioxide devices, functional integration with active photonic devices and complementary metal oxide semiconductor (CMOS) circuit, and fabrication process compatible with a mature CMOS manufacturing technology. One of the state of art of the silicon
photonic modulator engine that is very useful for quantum interference is MZI. A typical 2 $\times$ 2 MZI modulator cell consists of two 3 dB coupler and a dual-waveguide arm between
them. One of the arms has a phase shifter based on the change of refractive
index. Since the silicon has both strong thermo-optic (T-O) effect (1.86$\times$10$^{−4}$ K$^{−1}$)\cite{CMOSSensors} the phase shifter can be categorized as T-O switch with a heater and
electro-optic (E-O) switch with a p-i-n junction diode. The T-O switch has a response time of microsecond-scale to millisecond-scale, while the E-O switches have a response time of nanosecond-scale. 

The loss in on-chip optical interferometers arise from non-unity coupling from fiber to the input ports of the chips as well as attenuation through the waveguides patterned on the chip. As discussed above, the coupling loss can be significantly less than 1~dB  through the advanced coupling methods. However, the waveguide propagation loss is given by the chip platform. Depending on the wavelength, this loss can vary substantially, in particular in the wavelength range of 300-700~nm, as shown in Table~\ref{tab}.

\begin{figure*}
    \centering
    \includegraphics[scale=0.20]{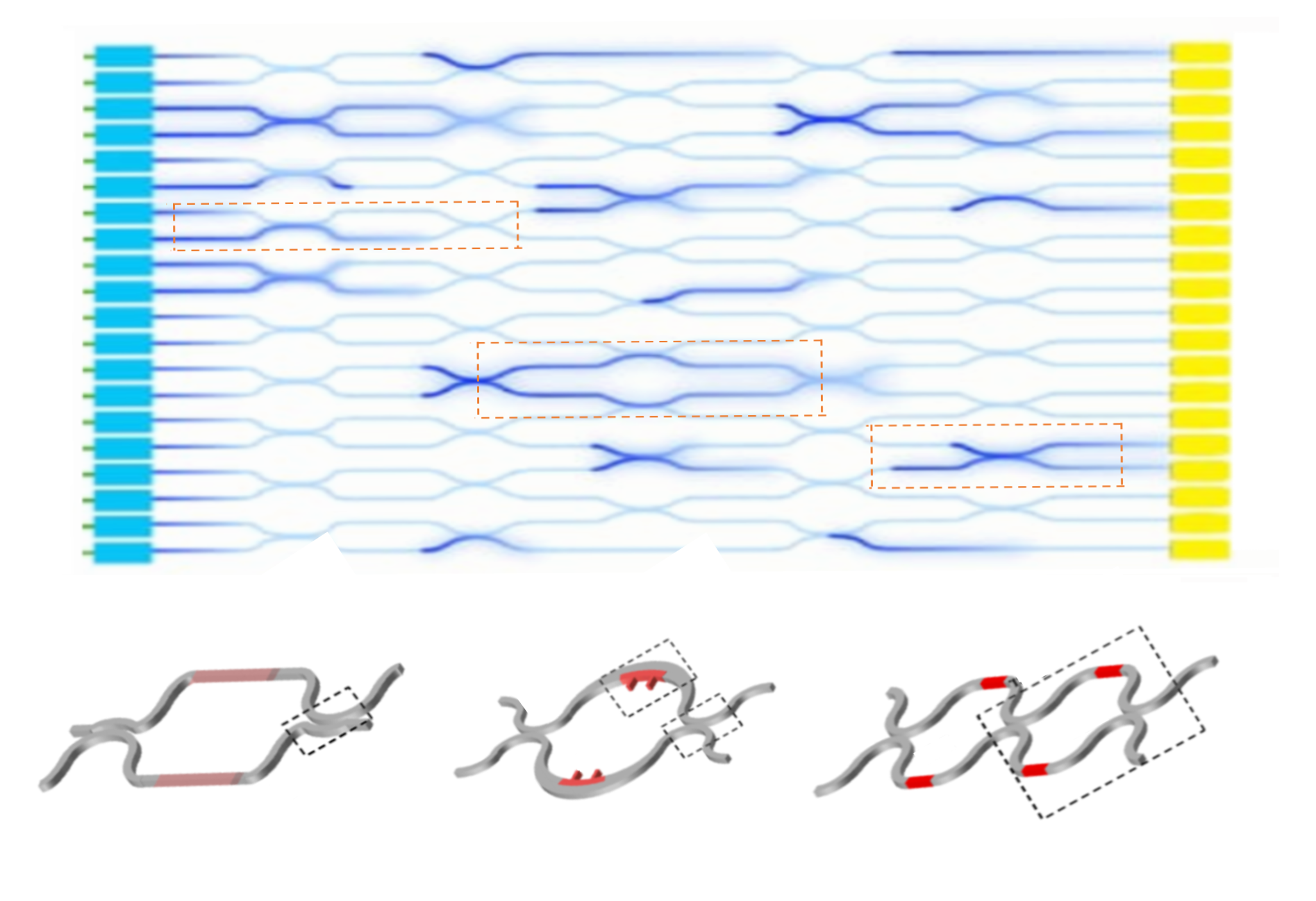}
    \caption{Schematic of various of MZI modulator cells in an optical interferometer. The undesirable attenuation of light in the waveguides and modulators depends on material of the chip platform as well as the dimension and structure of modulators\cite{CMOSSensors}, which determine bending and scattering loss.}
    \label{fig:6}
\end{figure*}

\subsection{Noise and Loss in the CMOS sensor array}
Noise can be produced by fluctuations in signal that makes uncertainty in detection. Essentially, the signal-to-noise ratio (SNR) is the ratio of pattern signal to the total noise. For larger SNR it is easier to distinguish pattern from noise, which makes a higher confidence in measurements.\\
CMOS (Complementary metal–oxide–semiconductor) primary noise sources are shot (photon) noise (i.e. SN), dark noise (i.e. DN), and read noise (i.e. RN). Shot noise is due to physical property of light, regardless of sensor, and it is $\text{SN}=\sqrt{\text{Signal}}$. Dark noise is temperature dependent and higher for global shutter and its magnitude is obtained as
$\text{DN}=\sqrt{\text{Dark}\;\text{Current}}$. Read noise includes Random Telegraph Noise (RTN), which is non-Gaussian, and depends on multiply column and pixel amplifiers, $\text{RN}=\text{Read\;Noise}$. RTN is the most significant component of CMOS noise. The SNR for CMOS is obtained as follows

\begin{equation}
    \text{SNR}=\frac{S}{\sqrt{{\text{SN}}^2+{\text{DN}}^2+{\text{RN}}^2}}
\end{equation}
where $S$ is Signal$=$Photon flux $\times$ time $\times$ quantum efficiency (QE) \cite{QECMOS}.

Scientific CMOS (sCMOS) sensor is a novel technology with room to grow, which allows for higher speed operation with larger pixel arrays than EMCCD and CCDs with similar noise performance to conventional CCDs. 

\subsubsection{Quantum Efficiency}
Quantum efficiency (QE) is defined as

\begin{equation*}
    \text{QE}=\frac{\text{Converted photons to electrons}}{\text{Total incident photons}}
\end{equation*}
which is a measurement of sensitivity to light. As a ratio, QE is dimensionless, but it is closely related to the responsivity, which is expressed in amps ($A$) per watt ($W$). Since the energy of a photon is inversely proportional to its wavelength, QE is often measured over a range of different wavelengths to characterize a detector efficiency at each photon energy level.

The photodetector matrix consist of CMOS-compatible photodiodes (formed between drain diffusion and p-well) with associated readout and sensor selection circuits. The spectral measurements of the photodiode have exhibited a quantum efficiency better than 60\% at 650 nm, and better than 40\% between 500 and 850nm \cite{QECMOS}.\\
A chip design for UPE detection can be inspired by retina implants, but with bigger array size and significantly higher quantum efficiency. The irradiance on the retina even under a bright daytime illumination does not exceed 1 $\mu$W mm$^{-2}$. At such illumination a 20 $\mu$m diameter photodiode (having even 100\% quantum efficiency) can provide only 40 pA of current \cite{Retinaimplant}. Basically, each photoreceptor cell can produce 1 pA with a single photon absorption \cite{SalariPlosone}. 
To provide stimulating current on the order of 1-2 $\mu$A, which would be minimal for physiological stimulation, current amplification by a factor of about 1000 is required. Suitable current levels would require photodiodes more than 600 $\mu$m in diameter, so that ambient light cannot be used to power more than a token number of electrodes on a retinal chip. An additional source of power will be needed for any practical chip \cite{Retinaimplant}.  The stimulation current for an electrode of 10 $\mu$m in diameter is on the order of 1 $\mu$A. The photodiode converts photons into electric current with efficiency of up to 0.6 AW$^{-1}$, thus 1.7 $\mu$W of light power will be required for activation of one pixel. If light pulses are applied for 1 ms at 50 Hz, the average power will be reduced to 83 nW/pixel. With 18000 pixels on the chip, the total light power irradiating an implant will be 1.5 mW \cite{Retinaimplant}. In the case of skull-implant PIC chip, the main difference with the retina implant is that the retina implant should activate neurons with the currents produced by external light, which needs a relatively high intensity of light, while for the PIC chip there is no need to activate neurons, and a low light intensity even with a few numbers of photons is sufficient for the CMOS pixels activation to be reported to the software. In silicon, a single-photon with a wavelength between 300 and 1100 nm can generate only one electron–hole pair. Therefore, for visible and near-infrared light, the task of single-photon detection becomes a task of single-electron (or hole) detection. This is not easy due to the unavoidable readout noise of the sensor, which is usually too high for the reliable detection of a single electron. Another difficulty for room temperature applications are the thermal dark currents, because they are indistinguishable from photogenerated signals.

\subsubsection{Chip battery and Wireless sectors}
In order to have a dynamic chip for monitoring signals of the brain continuously, the chip requires a long lifetime battery. The size and lifetime of the battery is one of the major challenges in design of an implant chip for biomedical applications. As an alternative, replacing the battery with a miniaturized and integrated wireless power harvester aid the design of sustainable biomedical implants in smaller volumes \cite{Antenna}.
Currently, implanted batteries provide the energy for implantable
biomedical devices. However, batteries have fixed energy density, limited lifetime,
chemical side effects, and large size. Thus, researchers have developed several methods
to harvest energy for implantable devices. Devices powered by harvested energy have
longer lifetime and provide more comfort and safety than conventional devices. A
solution to energy problems in wireless sensors is to scavenge energy from the ambient
environment. Energies that may be scavenged include infrared radiant energy, wireless transfer energy, and RF radiation energy (inductive and capacitive coupling) \cite{Hannan}. 
Recently, a chip has been developed that is powered wirelessly and can be surgically implanted to read neural signals and stimulate the brain with both light and electrical current. The technology has been demonstrated successfully in rats and is designed for use as a research tool. The chip is capable of 16-ch neural recording, 8-ch electrical stimulation, and 16-ch optical stimulation, all integrated on a 5 $\times$ 3 mm$^2$ chip fabricated in 0.35-$\mu$m standard CMOS process. The trimodal SoC is designed to be inductively powered and communicated \cite{Wirelesschip}.

\subsection{Biocompatibility of the chip}
Brain implants may induce side effects; for instance they may interact acutely and chronically with the brain tissue possibly causing blood-brain barrier (BBB) breach, vascular damage, micromotions, diffusion etc \cite{Biocompatibility1}. The advantage of our suggested photonic chip is that it is minimally invasive compared to invasive implants (e.g. ECoG) since it does not need to penetrate the brain tissue.\\ Some of the key fundamental questions associated to brain implants are related to how long an implant can record useful neuronal signals and what degree of acquisition and decoding reliably can be achieved if the tissue is affected by chip implant. Functional neural tissue survival, distance from the chip contact to target and long term stability are essential parameters to be considered \cite{Biocompatibility1}.\\
In the case of photonic chip, it should be installed on the inner surface of the skull and not to be implanted directly in the brain tissue. However, there is still the possibility of a close contact with the brain meninges (i.e layered membranes that protect the brain and spine) due to the mechanical or volume changes of the brain. In this case, it has been shown that Silicone causes the least amount of inflammation relative to other materials tested at all sacrifice points, which makes it the leading standard neurosurgical implant material and an appropriate control for studies of brain biocompatibility \cite{Biocompatibility2}. Thus, we envisage to adopt Silicone chips but we also expect that research in biocompatibility will provide alternative and advanced materials. However, since the photonic chip can be implanted in between the meninges and the skull, there can be concerns about the limitation of brain UPE detection due to the existence of meninges. The meninges layers of the human brain are composed of three main layers: dura, arachnoid, and falx. The key question is whether light can pass through these layers and if it does, then what are the scattering and absorption effects of photons? For instance, to have a reasonable data acquisition should the dura be open? The optical properties of the human brain and its meninges have been investigated decades ago. It has been shown that meninges is approximately transparent for the near-IR range, but almost half of emissions will not pass through it in the visible range, and less than 40\% of emissions can pass through the meninges in the UV range (200-400nm) \cite{meninges}. As a result, based on the high efficiency of the photonic chip in the near-IR range, the existence of meninges reduces the intensity of UPE but it does not lead to a significant limitation.\\
Additionally, because of the aqueous and biochemically aggressive nature of the body, the lifetime of brain implants strongly depends on packaging. There are different methods for packaging, which may be especially important for the case of traditional electric chips with wireless neuromodulatory implants with increasing electrode count to have an in-vivo lifetime comparable to a sizable fraction of a healthy patient's lifetime ($>$10-20 years) \cite{packaging}. For our suggested photonic chip, the situation is considerably better because the chip does not have electrodes in the wet biological tissue nor contact with that, and the environment between the meninges and skull is not aqueous and therefore the probability of water leakage in the photonic chip is minimal. If there will be an injury in the meninges layers due to some impact or accident, then the aqueous leakage may occur, where the photonic chip should be investigated for packaging based on the materials are used in.

\section{Discussion and Conclusion}

We propose a radically novel brain-computer interface (BCI) that is based on ultraweak photon emission (UPE) from the brain. We describe its feasibility of fabrication based on integrated photonic circuits that be readily implemented in a lab. The envisaged BCI chip can be implanted on the interior surface of the skull to monitor in real-time UPE signals emanating from the cortex surface. The proposed chip is not only useful for BCI technology but also it can be used as a photonic sensor for imaging, spectroscopy and sensitive measurements at low light-levels in several applications from biological UPE to quantum optical processing~\cite{QPRP}. Although our proposed technology is, admittedly, at the level of conjecture, requiring comprehensive tests and investigations for verification, we still envision complementary features as well as certain advantages over established technologies, including ECoG. The inherent advantage of our proposed technology is that it is minimally invasive when compared to ECoG. Furthermore, there are certain side effects that may affect the quality of data acquisition over time in ECoG, whereas we expect a relatively stable long-term data acquisition in our proposed approach. In addition, if our suggested photonic chip-technology reaches a satisfactory detection performance based on our estimations,  we anticipate that it can feature some other advantages. For example, it may provide additional information about brain functioning, such as an approximately real-time imaging (in slightly longer timescales, e.g. each 15, 20, 30, 60 min, or so) and open the door to studying metabolism variations, variation of ROS production, delayed luminescence but also undertake novel and complementary studies on object visualization studies, sleep studies and neurodegenerative diseases \cite{BreakspeaEtAL, FulopEtAL}. Indeed, the emphasis of our conjectural paper is to develop a novel technology and methods that could provide complementary information to improve our understanding of brain activity with potential applications for BCI technologies.

Now, we would like to discuss the advantages and limitations of our proposed technology versus the current BCI methods. On-chip PICs offer advantages such as miniaturization, higher speed, low thermal effects, large integration capacity, and compatibility with existing processing flows that allow for high yield, volume manufacturing, and lower prices. In the case of UPE detection, there is no need for on-chip single-photon sources, which is one of the most difficult challenges in PICs for quantum computation and communication. In the suggested chip, single photons are produced naturally by metabolism in neurons and therefore a lower power with battery is needed for energy consumption on an implant PIC. Loss is low in NIR range (e.g. $2\times 10^{-6}$ dB/cm). In addition, photons are bosons, which don’t interact and crosstalk is minimal. An PIC for optical interferometery is efficient for the wavelengths typically in the near infrared range, 800nm-1650nm. This makes a limitation for detection of UPE photons which are in the visible range and the overlapped part to NIR, 400nm-800nm. For example, loss is high for the visible range (e.g. 0.6 dB/cm at 600 nm).  

Moreover, it may look that the single-photon detections on a CMOS array have a low quantum efficiency besides the dark current in room temperature, which may lose considerable amounts of UPE. Another concern may be that the output of CMOS is electrons, which are charged particles and fermions, and therefore electronic crosstalk is inherent. In fact, the CMOS quantum efficiency (QE) is about 75\%, which is about three times higher than the photo-multiplier tubes (PMTs) with QE about 25\%. The SNR of a PMT at room temperature to detect UPE photons is about 1 to 2, thus a cooling system is required to cool down the PMT sensor to enhance the SNR to reach 3 and higher. Obviously, there is no cooling system on an PIC chip, but in this case, the QE of the CMOS sensor can compensate the lack of a cooling system. For a simple estimation, assuming a 1cm$\times$1cm chip and considering the length of each CMOS pixel is 4$\mu$m, it is possible to have 2500 CMOS pixels as the output port on the chip, including 50$\times$50 pixels on the ROP. According the estimations in the main text, the amount of total photon loss from the receiver optical plane (ROP) to the output of the optical interferometer (OI) is about 50\%, and the QE of CMOS at the output of the OI is estimated to be 25\% in body temperature under the implant conditions on the skull to have a final SNR from 1 to 2. Consequently, it is estimated that only 10\% of incident photons can be safely recognized in the output and reported wirelessly to the software on a computer or smartphone. Considering 10-1000 incident photons per second received in the ROP under a cognitive task (e.g. an object visualization), there can be 1-100 photons per second efficiently detected in the output port, which are enough to have a relatively successful implant PIC chip for an acceptable pattern for UPE processing, where the size of the machine learning program is $N\times N$ sparse matrix, which is not a difficult task for a chip size number of pixels.

 To conclude, in this paper, we advance major conjectures regarding the relevance of UPE patterns and decision making as well as the feature extractions from UPE signals, which need to be experimentally verified. However, despite some probable limitations in chip fabrication and efficiency, it may be used for wireless BCI signal acquisition with several advantages versus traditional counterparts such as speed, size, minimally invasive, cheap, scalability, etc. This can be a potential step forward for real time brain imaging and biological information processing.

\section*{Conflict of Interest Statement}
The authors declare that the research was conducted in the absence of any commercial or financial relationships that could be construed as a potential conflict of interest.

\section*{Funding}
VS and SR are grateful for the financial support by the Spanish State Research Agency through BCAM Severo Ochoa excellence accreditation SEV-2017-0718 and BERC 2018-2021 program. SR is also grateful for project RTI2018-093860B-C21 funded by (AEI/FEDER, UE) with acronym “MathNEURO”. CS acknowledges NSERC Discovery Grant RGPIN-2020-03945. 

\section*{Acknowledgments}
Authors thank M. Aslani for illustration of some of the figures. VS and SR are very thankful for several helpful discussions with Jean-Bernard Bru at BCAM.


\vspace{0.6 EM}

\noindent\textbf{Competing interests:} 
The authors declare that they have no competing interests.
\\

\end{document}